\documentclass[a4paper,10pt]{article}
\usepackage{amssymb}

\usepackage[noadjust]{cite}
\usepackage[utf8]{inputenc}
\usepackage[english]{babel}
\usepackage[nottoc]{tocbibind}
\usepackage{mathptmx}
\usepackage{graphicx}
\usepackage{epstopdf}
\usepackage{textcomp,gensymb}

\title{On the fly control of high harmonic generation using a structured pump beam}

\usepackage{authblk}
\author[1]{Lilya Lobachinsky}
\author[1]{Liran Hareli}
\author[1]{Yaniv Eliezer}
\author[1]{Linor Michaeli}
\author[1,*]{Alon Bahabad}

\affil[1]{Department of Physical Electronics, Tel-Aviv University, Tel-Aviv 69978, Israel}
\affil[*]{alonb@eng.tau.ac.il}

\begin{document}
	
	\date{}
	\maketitle

	\begin{abstract}
		High harmonic generation (HHG) is an extreme nonlinear frequency up-conversion process during which extremely short duration optical pulses at very short wavelengths are emitted. A major concern of HHG is the small conversion efficiency at the single emitter level. Thus ensuring that the emission at different locations are emitted in phase is crucial. At high pump intensities it is impossible to phase match the radiation without reverting to ordered modulations of either the medium or the pump field itself, a technique known as Quasi-Phase-Matching (QPM). To date, demonstrated QPM techniques of HHG were usually complicated and/or lacked tunability. Here we demonstrate experimentally a relatively simple, highly and easily tunable QPM technique by using a structured pump beam made of the interference of different spatial optical modes. With this technique we demonstrate on-the-fly, tunable quasi-phase-matching of harmonic orders 25 to 39 with up to 30 fold enhancement of the emission.

	\end{abstract}

	\section*{Introduction}
	High Harmonic Generation (HHG) is an extreme non linear optical up-conversion process driven with an intense ultra-short laser pulse, usually interacting with noble gas atoms  \cite{brabec2000intense,kapteyn2007harnessing,corkum2007attosecond}. The up-converted light can easily reach the extreme UV portion of the spectrum and in certain cases can extend up to keV  x-ray photon energies \cite{popmintchev2012bright}.   At the single atom level the process of HHG  is most simply
	described by the three step model \cite{corkum1993plasma,lewenstein1994theory} consisting of electron ionization, laser-driven acceleration and recombination upon which excess kinetic energy is released in the form of a high energy photon.

	Macroscopically, due to dispersion and geometric effects such as wave-guiding and focusing, the efficient build up of each harmonic order is limited to its coherence length $L_c=\pi / \Delta k_q$  which is the distance at which the accumulated phase difference between the pump beam and the harmonic emission is equal to $\pi$. Here, $\Delta k_q$ is the momentum phase mismatch -  the difference between the waves vectors of the pump beam and the q order harmonic signal at the direction of propagation.

	Perfect phase matching conditions, $\Delta k_q = 0$, can be achieved at low intensities (low ionization rates) and loose focusing \cite{durfee1999phase,popmintchev2008extended}. However, when working with a strong pump beam, plasma dispersion becomes dominant and phase matching is not possible \cite{popmintchev2009phase}. Thus, a different approach is required for achieving efficient build up of the emitted radiation. The solution comes in the form of an ordered modulation of a parameter relevant to the interaction, which allows the restrictive momentum conservation condition associated with perfect phase matching to be replaced with a less restrictive condition \cite{lifshitz2005photonic} known as quasi-phase-matching (QPM) \cite{armstrong1962interactions,bahabad2010quasi} . In particular using a periodic spatial modulation allows quasi-phase-matching when $\Delta k_q=2\pi / \Lambda$, where $\Lambda$ is the period of the modulation.   
	
	To date, most of the works applying QPM on HHG, did this by modulating the medium properties\cite{dromey2007quasi,ganeev2014quasi,gibson2003coherent,pirri2008enhancing,paul2003quasi,ganeev2015influence} while others showed all optical QPM theoretically \cite{kovacs2012quasi,serrat2010all} and experimentally \cite{zhang2007quasi,o2014quasi}. While some of these works demonstrated tunability they were associated with mechanically moving components. In addition, although a relatively simple method for a co-propagating all optical QPM was suggested \cite{faccio2010modulated}, the all-optical methods that were demonstrated experimentally involved counter-propagating geometries which led to relatively complicated front ends \cite{o2014quasi,zhang2007quasi,kapteyn2007harnessing}.

	Here we present an all-optical co-propagating scheme for QPM of HHG in a semi-infinite gas cell configuration, allowing for non-mechanical, on-the-fly and complete control of the parameters of a perturbing periodic standing-wave modulation. In particular, the period of the modulation, its overall length and the depth of the modulation are easily controlled. With this scheme we experimentally demonstrate tunable QPM of harmonic orders 25 to 39, with enhancement of up to 30 fold.   
	
	\section*{Results}
	\subsection*{Theory}
	
	HHG in a so called semi-infinite gas cell \cite{peatross2004phase} is driven with a pump beam which is focused to a point at the vicinity of a small exit hole in a thin foil terminating a gas cell. Vacuum conditions are established immediately after the cell. The phase mismatch for the generation of the q-th harmonic is given by: \cite{rundquist1998phase,auguste2007quasi,balcou1997generalized}  
	
	\begin{eqnarray}	
		\Delta k_q &=& qk_\omega-k_{q\omega}\simeq\nonumber\\&\simeq&qP\left((1-\eta)\frac{2\pi}{\lambda} \Delta n_q-\eta N_{atm} r_e \lambda \right)+q\partial_z \phi_{geometric}+\partial_z \phi_q, 
		\label{eq:Deltak}
	\end{eqnarray}

	where $q$ is the harmonic order, $k_{\omega}$ is the pump wave vector, $k_{q\omega}$ is the q-th harmonic wave vector,  $P$ is the gas pressure, $\eta$ is the ionization rate at time of emission of the harmonic radiation, $\lambda$ is the pump wavelength, $\Delta n_q$ is the difference of refractive indices of neutral gas atoms between the pump and the harmonic radiation, $N_{atm}$ is the number density of atoms at atmospheric pressure, $r_e$ is the classic electron radius, $\phi_{geometric}$ is the geometric phase and finally $\phi_q$ is the intrinsic phase of the atomic dipole which is proportional to the pump intensity  \cite{lewenstein1995phase,balcou1997generalized}.
	
	The two pressure-dependent terms in Eq.\ref{eq:Deltak} are due to gas and  plasma dispersion respectively. The third term is the geometric wave vector mismatch which is usually approximated to depend only on the geometric variation of the pump phase and is mostly associated  with either wave guiding effects or with focusing in free propagation (contributing a Gouy phase term). In the present case, the use of  a structured pump beam made of the interference of a Gaussian and a Bessel beam would modulate both the geometric and the intrinsic phase terms.

\begin{figure}[ht]
		\centerline{\includegraphics[width=1\linewidth]{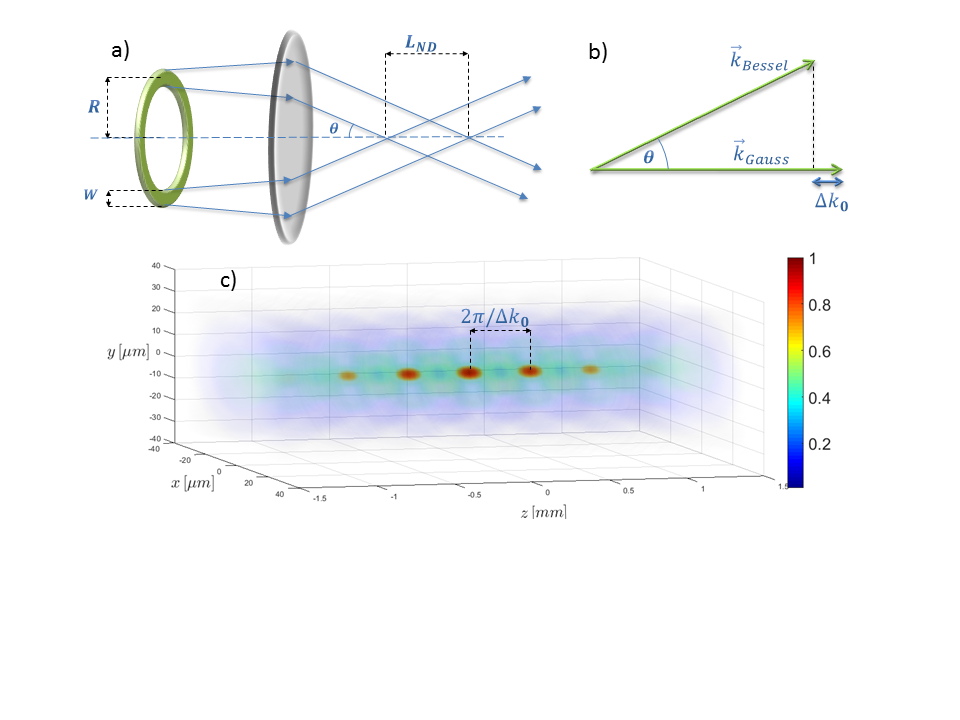}}
		\caption{ \textbf{Creating a periodic intensity and phase modulation using an interference of a Gaussian and a Bessel beam.} (a) Schematic representation of a Bessel beam which is non diffracting along $L_{ND}$, generated by focusing a ring amplitude field distribution in a 2f geometry.
			(b) The wave vectors of the interfering Gaussian beam $k_{Gauss}$ and the Bessel beam $k_{Bessel}$. The on-axis difference between the wave vectors is $\Delta k_0$. (c) A volumetric representation of the intensity of the superposed Gaussian and Bessel beams}
		\label{fig:deltak}
	\end{figure}

	The spatial spectral decomposition of a Bessel beam \cite{mcgloin2005bessel} is made of a continuum of plane waves aligned on the surface of a cone with a half angle $\theta$ with respect to the propagation axis. 
	The difference between the Gaussian wave vector $k_{Gauss}$ and the on-axis projection $k_{Bessel}$  of the Bessel wave vector, is denoted with $\Delta k_0$ (see Fig.\ref{fig:deltak}.(b)). When the two beams are superposed coaxially to produce a structured pump beam, a periodic intensity and phase modulation, with an underlying period of ${2\pi}/{\Delta k_0}$, emerges on axis (see Fig.\ref{fig:deltak}.(c)).

	The on-axis intensity and geometric phase of the structured pump beam can be roughly estimated by: 
	
	\begin{eqnarray}			
		\phi(z)_{geometric}&=&\int_{0}^{z} k(z)_{geometric}dz=\phi_{Gouy}(z)-atan \left(\frac{\beta sin(\Delta k_0 z)}{1+\beta cos(\Delta k_0 z)}\right)sinc(\frac{\pi z}{L_{ND}}) \nonumber	
		\\
		I_{total}(z)&=&I_{Gauss}(z)\left( 1+2\beta cos(\Delta k_0z)sinc(\frac{\pi z}{L_{ND}})\right)
		\label{eq:GeometricModulation}
	\end{eqnarray}
	where $\beta$ is the amplitude modulation depth (amplitude ratio between the Bessel and Gaussian fields at the focus of the Bessel beam), $\phi_{Gouy}(z)=atan(z/z_R)$ is the on-axis Gouy phase of the Gaussian beam where $z_R$ is the Gaussian beam Rayleigh range, $I_{total}(z)$ is the total intensity of the superimposed beams, $I_{Gauss}(z)$ is the intensity of the Gaussian beam and  $L_{ND}$ is the non-diffracting distance which is defined later in Eq.\ref{eq:L_{ND}}. The intensity modulation acts on both the amplitude and on the phase components of the nonlinear polarization which has a tendency to oppose each other for  emission buildup associated with QPM \cite{diskin2015phase,hadas2016periodic}. In the current case, taking all modulation factors into account, the modulation depth of the amplitude and of the phase of the nonlinear polarization is different, thus the overall modulation can be used, as we verify here, to realize an efficient all-optical co-propagating QPM scheme.
	
	In order to produce a superposition of a Gaussian and a Bessel beams at the interaction region, we split a Gaussian beam to two and modulate the spatial distribution of one part using a Spatial Light Modulator (SLM).  The modulated beam obtains the shape of a ring which is a Fourier transform of a Bessel beam \cite{mcgloin2005bessel, ruffner2012optical}. Focusing the ring in a 2f configuration creates a Bessel beam which is non-diffracting along a distance $L_{ND}$ around the focus of the lens (see Fig.\ref{fig:deltak}.(a)). Using simple geometrical optics it can be shown that $L_{ND}$ is dependent on the ring thickness $W$, radius $R$ and the lens focal length $f$:
	\begin{eqnarray}	
		L_{ND}&=&	\frac{2\lambda f^2}{WR}
		\label{eq:L_{ND}}
	\end{eqnarray}
	
	When the structured beam is used as a pump for HHG, quasi phase matching for the q-th harmonic order can be achieved when $\Delta k_0=\Delta k_q$. For a given focal length $f$, the radius of the imaged ring, R, determines the angle $\theta$ of the Bessel wave vectors and consequently determines the value of $\Delta k_0$:
	
	\begin{eqnarray}	
		\Lambda&=&\frac{2\pi}{\Delta k_0} =\frac{\lambda/2}{1-cos(\theta)}
		\label{eq:Lp}
	\end{eqnarray}

	The number of periods within the modulation is given with $L_{ND}/\Lambda$ where $L_{ND}$ in turn is dependent on the thickness of the imaged ring (see Eq.\ref{eq:L_{ND}}). Thus $L_{ND}$ determines the effective phase-matching bandwidth $\delta k$ which roughly behaves as $\delta k=\pi/L_{ND}$. This follows from approximating the intensity buildup of the harmonic emission for a given modulation depth as proportional to $L_{ND}^2 sinc^2((\Delta k-\Delta k_0) L_{ND}/2)$ \cite{boyd2003nonlinear} with $\Delta k$ being any given value of the phase mismatch. However, it is important to note, that for an efficient use of the available number of periods (as well as for the last approximation to be of any value) it is essential that the effective interaction length, determined by the Gaussian Rayleigh range, is at least as long as $L_{ND}$. This is easily accomplished by using a shallower focusing for the Gaussian beam than for the Bessel beam. The depth of the modulation is determined by the ratio of the intensities of the two beams. Using thinner rings increases $L_{ND}$ but decreases the intensity of the Bessel beam and so reduces the modulation depth. 
	Another factor that modifies the modulation depth is the ring radius -  due to the Gaussian  profile of the beam reflecting of the SLM, larger ring radius would decrease the intensity of the Bessel beam.  Finally, the modulation depth can also be tuned by adjusting the overall transmission of any given ring, which is also controllable with the SLM. In our experiment the rings transmission was set to its maximum value.      
	
	As the radius and width of the rings are controlled by the software operating the SLM it is very easy to exert on-the-fly control on the geometric parameters of the quasi-phase-matching modulation - effectively changing the center of the phase-matching curve $\Delta k_0$, its bandwidth $\delta k$ and its modulation depth.
	
	\subsection*{Experiment}
	In the experiment an intense Gaussian beam and a perturbing Bessel beam are combined and focused together in a Semi-Infinite Gas Cell (SIGC). The exact parameters of the Bessel beam  are easily determined using a computer-controlled beam shaper based on a phase-only Spatial Light Modulator (SLM) set in amplitude configuration (see Methods). In the interaction region the beams interfere and form a periodic standing wave pattern (see Fig.\ref{fig:deltak}.(c)). 
	
	The Rayleigh range of the Gaussian beam that was used is 5mm. The non-diffracting Bessel region $L_{ND}$  varies between 1.4mm and 8.3mm while the periodicity of the induced periodic modulation varies from 200$\mu$m to 430$\mu$m. Thus the range of the number of periods in the modulation can vary between 3 to 41. These ranges are determined by the SLM area and by the imaging optics being used in the setup.

	\begin{figure}[ht]
		\centerline{\includegraphics[width=1\linewidth]{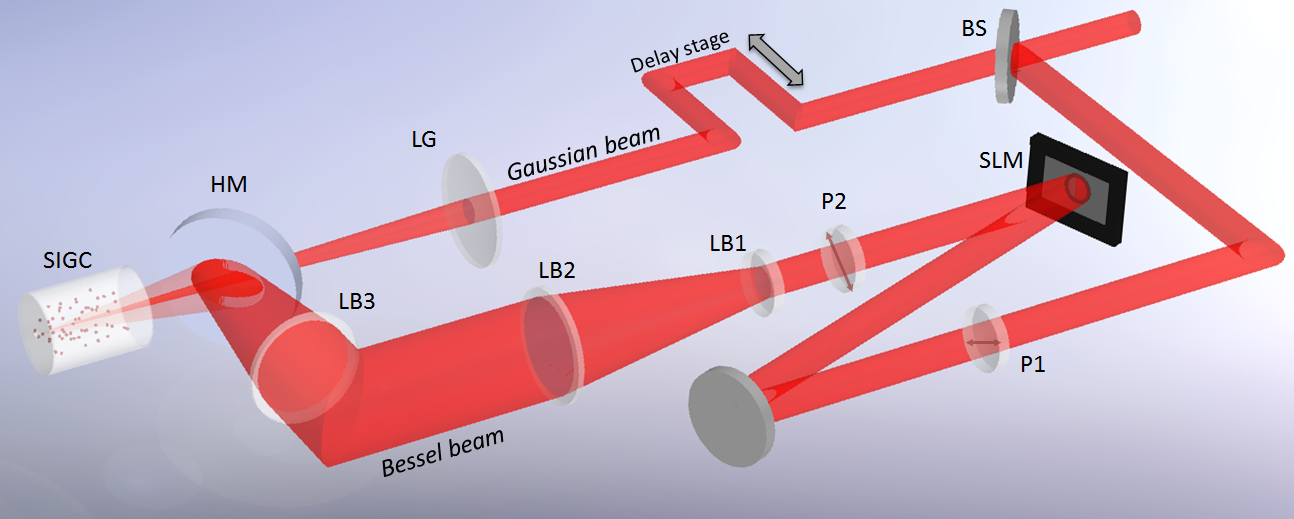}}
		\caption{Schematic representation of the experimental setup, where LB1-3 are lenses along the Bessel beam path, LG is the Gauss beam lens, BS - Beam-splitter, P1-2 - Polarizers, SLM - Spatial Light Modulator, HM - Holed mirror and SIGC - Semi-infinite gas cell. The delay stage allows to compensate for the difference between the optical paths of the two beams, assuring time coincidence of the two pulses at the interaction region.}
		\label{fig:Setup}
	\end{figure}

\begin{figure}[ht]
		\centerline{\includegraphics[width=1.2\linewidth]{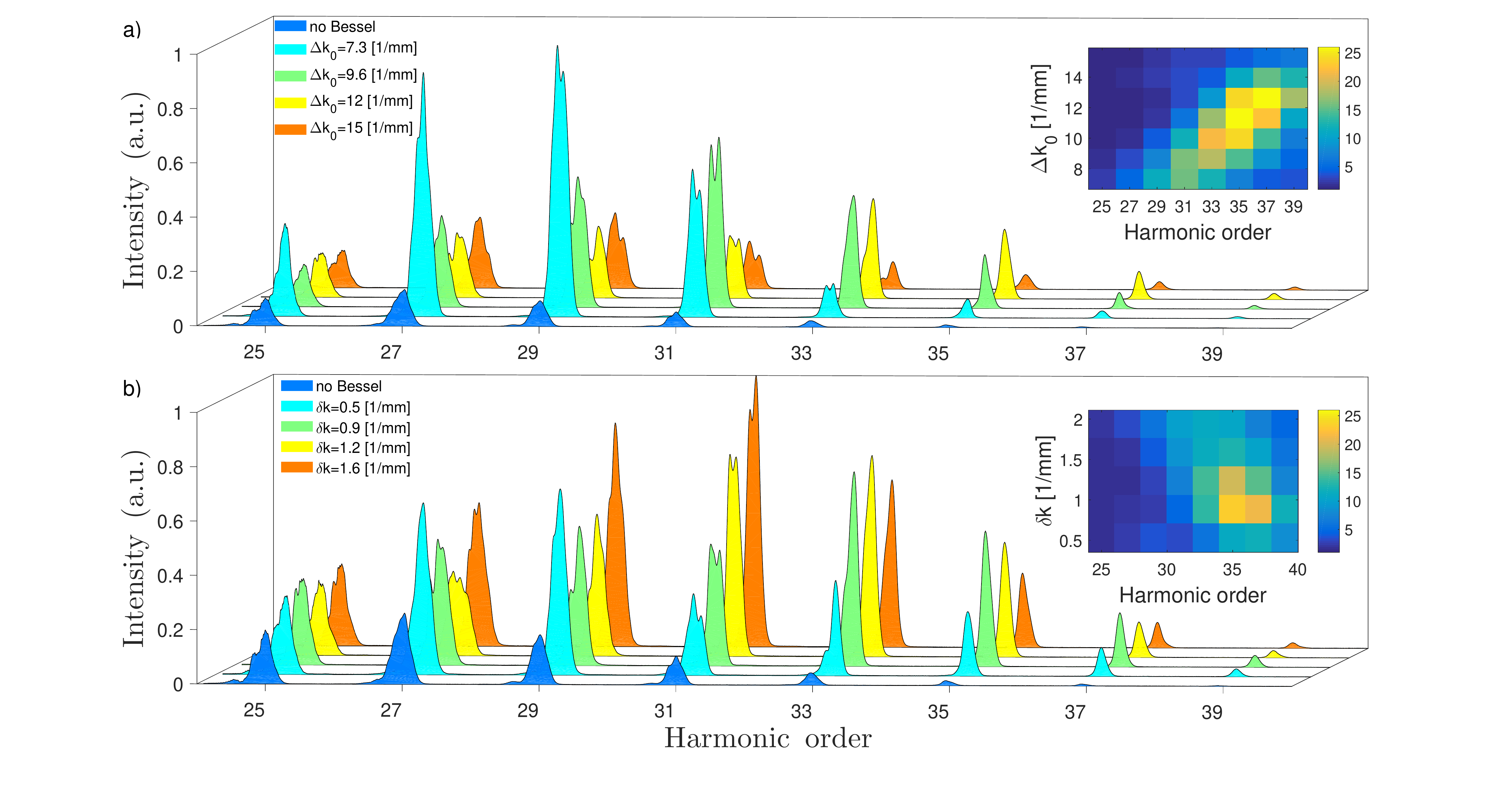}}
		\caption{ Harmonic enhancement: (a) Harmonic spectra produced by a Gauss beam only (blue) and with a modulation caused by an addition of a co-propagating Bessel beam with different  $\Delta k_0$ (other colors) for a fixed phase matching bandwidth of $\delta k=0.83 \pm 0.16 mm^{-1}$. Inset: The integrated intensity enhancement of each harmonic order for different  $\Delta k_0$.  
			(b) Harmonic spectra produced by a Gauss beam only (blue) and with a modulation caused by an addition of a co-propagating Bessel beam with different $\delta k$ (other colors). Inset: The integrated intensity enhancement of each harmonic order for different $\delta k$ . In this case the radius of the Bessel ring  corresponded to $\Delta k_0=12 mm^{-1}. $}
		\label{fig:Results}
	\end{figure}

	The HHG spectrum for harmonic orders 25 to 39 when only the Gaussian beam is present and focused to 1.5mm before the exit hole of the SIGC, is shown in Fig.\ref{fig:Results} (in blue). At the absence of the modulation due to the interference with the Bessel beam, the harmonic orders are not phase matched and therefore do not build-up efficiently. The addition of a Bessel beam with a specific cone angle for generating a modulation with the appropriate periodicity needed to phase match the harmonic orders under scrutiny, led to a substantial enhancement of these harmonics. Some of the harmonic orders that were nearly undetected before the addition of the Bessel, due to poor phase matching condition, became significant.

	The conditions at which the harmonic radiation is most efficient is a complicated function of the time-dependent intensity: higher intensity yields a stronger dipole moment, however it also increases the ionization rate which changes the phase mismatch. 
	We assume that the applied periodic modulation which provides the highest enhancement of the harmonic radiation matches the actual phase mismatch. For example, applying a modulation of $\Delta k_0=10 mm^{-1}$ the strongest enhancement was observed for the 33 harmonic therefor we conclude that $\Delta k_{33}=10 mm^{-1}$. Interestingly, this corresponds to ionization rate of $0.3$ when using  Eq.\ref{eq:Deltak}, while the maximum ionization rate is 0.81 (see Methods). This suggests that the best conditions for the generation of this harmonic order are at the leading edge of the pulse and not at its maximum. 
	
	Increasing the frequency of the modulation $\Delta k_0$, by modifying the radius of the ring on the SLM (for a fixed ring width, in which case the phase matching bandwidth is $\delta k=0.83 \pm 0.16 mm^{-1}$) we clearly observe a displacement of the band of phase-matched harmonics to higher orders. This is easily seen for the integrated intensity enhancement shown in the inset in Fig.\ref{fig:Results} (a). The enhancement was calculated by dividing the integrated intensity of each harmonic order by the integrated intensity of the same harmonic order produced by the unperturbed Gaussian beam. The observed displacement is in accordance with the expected linear dependence of the phase mismatch on the harmonic order (see Eq.\ref{eq:Deltak}). This verifies that the observed enhancement is due to quasi-phase-matching (furthermore, we remind that the intensity of the perturbing Bessel beam is about two orders of magnitude lower than that of the Gaussian beam and so the added intensity alone cannot account for any observed enhancement).  
	Overall the most enhanced harmonic order is scanned from the 27th to the 37rd harmonic and enhancement of up to 23 times can be observed for this particular case of relative positions of the focus of both beams and the exit hole of the SIGC.
	Next we modify the phase-matching bandwidth $\delta k$ by modifying the ring thickness on the SLM, while keeping a constant value of $\Delta k_0=12 mm^{-1}$. The results are shown in Fig.\ref{fig:Results}(b) with the integrated enhancement presented in the inset. Theoretically, according to the simplest QPM models \cite{boyd2003nonlinear}, the  phase matching bandwidth should grow linearly with $\delta k$. Analyzing the results we find that the linear fit of the phase matching bandwidth to $\delta k$ in our case has an $R^2$ fitting parameter of 0.81. In addition we see a redshift of the phase matched bandwidth which is not accounted for by simple models.

	\begin{figure}[ht]
		\centerline{\includegraphics[width=1.2\linewidth]{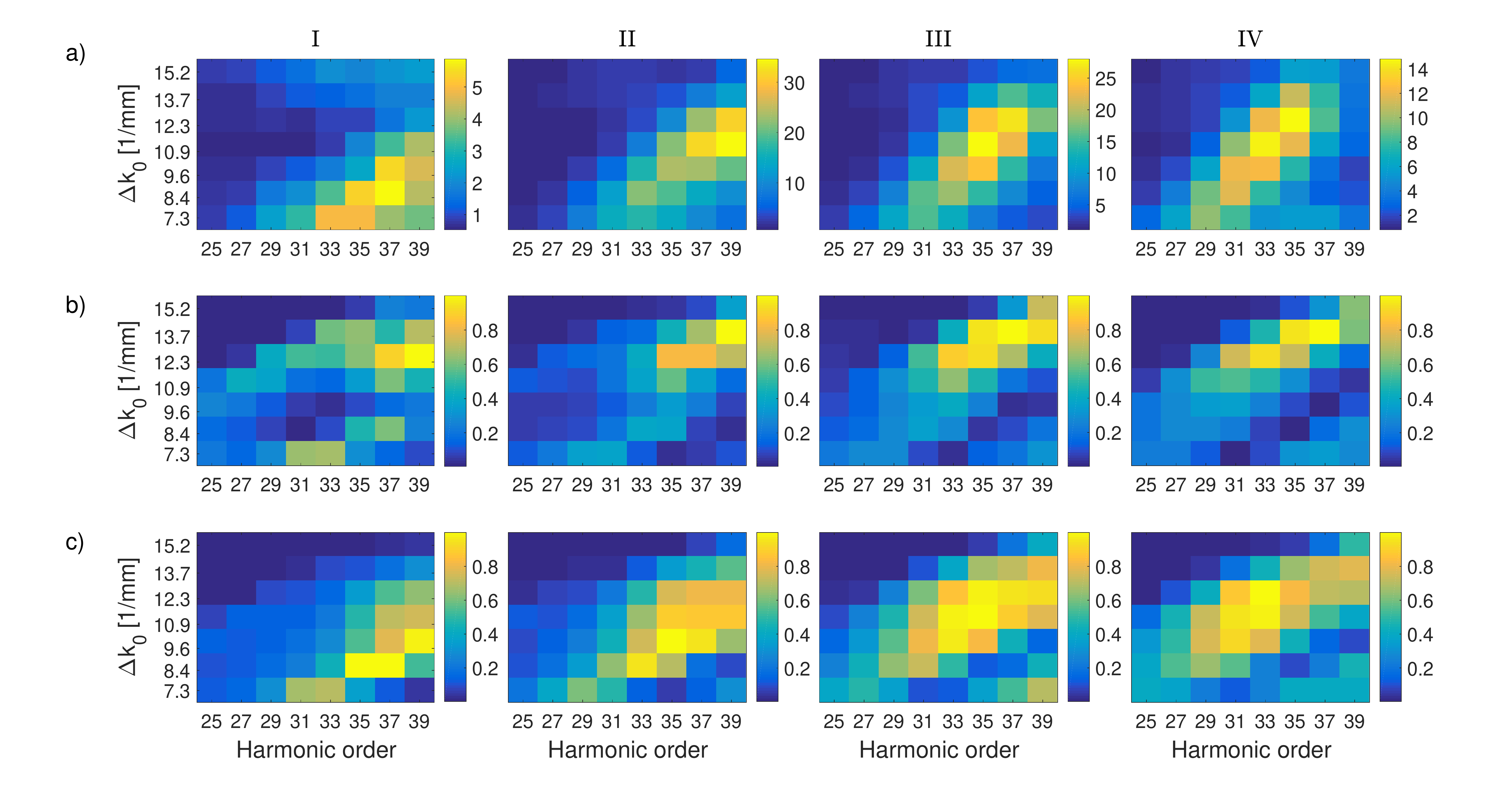}}
		\caption{HHG enhancement at different locations relative to the focus of the Gaussian beam. Column I to IV  are for the focus of the Gaussian beam being at 3.5 mm to 0.5mm before the SIGC exit at 1mm steps. At each focal location the modulation frequency $\Delta k_0$ is varied and the integrated harmonic enhancement is calculated.  In all cases the focus of the Bessel beam is fixed at 1mm before the SIGC exit. (a) Experimental results. (b) Numerical results (normalized). Modulation depth  $0.1$ of the experiment (c)  Numerical results (normalized). Modulation depth $0.3$ of the experiment. Neglecting geometric phase modulation.}
		\label{fig:diffRdiffpos}
	\end{figure}
	
	The above set of measurements was repeated for several different locations of the focus of the Gaussian beam with respect to the output of the SIGC, while keeping the position of the Bessel focus constant. By doing so, the QPM process takes place at different positions with respect to the Gaussian focus. The measured harmonics enhancement for the different cases is shown in Fig.\ref{fig:diffRdiffpos}(a). The highest observed enhancement in this set of measurements is 30 fold. But more importantly - assuming that the frequency of the modulation upon phase matching is close to the phase mismatch without the modulation, it is obvious that in all cases the phase mismatch depends linearly on the harmonic order: $\Delta k_0=Aq+B$ with both $A$ and $B$ dependent on the position of the focus of the Gaussian beam. The dependency of the $A$ and $B$ coefficients on the location upon the Gaussian beam profile can easily be explained using Eq.\ref{eq:Deltak}. The phase mismatch terms that are influenced by the changing of the location upon the Gaussian beam profile are the third and fourth terms (assuming that the pressure drops significantly only after the exit hole of the SIGC). The third term changes the proportionality factor in $q$ and so we can observe an increase in $A$ as the location gets nearer the focus of the Gaussian beam. At the same time, the fourth term increases together with the gradient of the intensity profile - increasing B as a consequence.  
	
	\subsection*{Simulations}
	
	A basic, even simplistic, simulation (see Methods) can capture the salient features of the observed results. We have fixed the geometric parameters of the simulations (focal location of the beams) according to the experiment. We explored a vast range of parameter space relating to the modulation depth of the perturbation, the ionization level, the intensity-proportionality factor of the intrinsic atomic phase, the ambient pressure, and the pressure gradient around the SIGC exit. In all cases where the perturbation is small enough the same general behavior is observed as in the experiment. However we did had to choose a  much smaller perturbation than estimated in the experiment or neglect the effect of the QPM modulation on the geometric phase (see Fig.\ref{fig:diffRdiffpos} (b)-(c)) to get similar results to the experiment. Otherwise the numerical result are much noisier.  The form of the perturbation of both the modulated intrinsic phase (proportional to the intensity) and of the modulated geometric phase are similar in form (behave as a sine function) while the main difference between the two - is that the intrinsic phase sine-like modulation sits upon a Gaussian background and the geometric phase upon an arctangent background (The Gouy phase), Our simulations indicate that, for the 1D simplistic model, the combination of the sine-like modulation and the Gouy background together is the main source for the deviation from the experimental results.    
	A plausible explanation for the difference between the parameters of the experiment and the simulation (to yield similar result) is that experimentally we gather harmonic radiation integrated over some effective transverse area of an interaction which is missing in the 1D model. In addition, the temporal dynamics (that is - the dependency of the dipole moment and of the phase mismatch on time) are missing in the simplistic numerical model we used and  their effect on the phase matching dynamics can be quite involved \cite{sandhu2006generation,thomann2009characterizing, bahabad2011manipulating}. Still, in both the experiment and simulations the linear behavior of $\Delta k_q$ as a function of $q$ is apparent, while the behavior of the offset parameter $B$ as a function of the Gaussian focus between the two is very similar.

	\section*{Conclusions}
	We have demonstrated experimentally driving of HHG using a structured pump beam made of the superposition of a Gaussian and a tunable perturbing Bessel beam. The structured beam is able to perform on-the-fly, tunable quasi-phase-matching of the harmonic spectra - with controlled band and bandwidth. We have demonstrated QPM of harmonics generated at different locations upon the main Gaussian beam. The flexibility of this method and its relatively easy implementation would allow to use a more complicated structured beams in order to control further parameters of the emitted radiation. For example - use of aperiodic modulations could allow for phase
	matching  several different bandwidths. It would also be very interesting to extend the use of structured pump beams to control other degrees of freedom of the harmonic emission, such as polarization states \cite{fleischer2014spin} and orbital angular momentum \cite{gariepy2014creating,vieira2016high}.  Additionally, perturbedly structured beams might be helpful to analyze in-situ conditions (such as ionization level, coherence length etc.) at different locations during the process of HHG. Finally, the all-optical technique presented here might also be applicable to perturbative nonlinear optical frequency conversion in nonlinear crystals \cite{bahabad2008quasi}.

	\section*{Methods}	 
	\subsection*{Experimental}
	
	To attain a relatively long Rayleigh range for the Gaussian  beam, and significant cone half angles $\theta$ for the Bessel beam, different focusing conditions are required for the two beams. As such, the output beam of a 1KHz, 35fs Ti:sapphire amplifier (Coherent Legend USX) is split into two paths (see Fig.\ref{fig:Setup}). The first beam having 0.4mJ per pulse, retains its spatial Gaussian  profile, and is used as the main pump beam. This intense beam also drills a through hole at the aluminum foil terminating the gas cell (in a  Semi-Infinite Gas Cell (SIGC) configuration). The second beam with a typical peak intensity which is only about 1.5$\% $ of the first beam (when material dispersion and achromatic pulse front tilt due to focusing are taken into account \cite{weiner2011ultrafast}), undergoes an amplitude modulation using two perpendicularly oriented polarizers and a phase-only Spatial-Light-Modulator (SLM,  Holoeye Pluto) in-between. This beam acquires a spatial distribution of a ring which is then imaged and focused to form a Bessel beam (see Fig.\ref{fig:deltak}.(a)). Both beams are combined using a holed mirror and are focused close to the output of the SIGC. The use of a holed mirror is possible because the Bessel beam retain its ring shape before the focus. In the interaction region the beams interfere and form a periodic standing wave pattern (see Fig.\ref{fig:deltak}.(c)). Modifying the position of both focusing lenses (LG and LB3 in Fig.\ref{fig:Setup}) allows for a careful selection of both the location of the interaction region with respect to the aluminum foil and the relative focus of both beams.
	The backing pressure of the SIGC was 15 Torr. The power of the Gaussian beam at the interaction region was 420 mW corresponding to a maximum ionization rate of 0.81 and to the HHG cutoff to correspond to the 187 harmonic order. The harmonic orders observed in the experiment are far below the cut-off. 
	
	\subsection*{Numerical}
	We have calculated using standard Fresnel propagation the on-axis spatial component of the structured beam. Then we approximated the behavior of the HHG field by integrating for each harmonic order the equation $dE_q/dz \propto N(I_{total}(z))^p exp(i \int{\Delta k_q dz})-\frac{\alpha(\lambda)}{2}E_q$ where $N$ is the density number of emitters and is proportional to the pressure $P$, $\alpha(\lambda)$ is the pressure-dependent power absorption coefficient of the $q$-th harmonic with $\Delta k_q$ calculated using Eq.\ref{eq:Deltak} where the geometric phase is taken from the pump field propagation calculation. The gas pressure gradient in the interaction region was calculated according to \cite{sharipov2004numerical}. We chose the power factor $p=5/2$ \cite{l1991higher}.  We assumed that only short trajectories are contributing to the emission by choosing the proportionality factor of the intrinsic atomic phase to the field intensity to be $\alpha_q=2\times 10^{-14} rad \cdot cm^2/W$ (its actual value varies among various models \cite{auguste2007quasi,gaarde2008macroscopic,rego2016nonperturbative}).

	\section*{Acknowledgments }	
	This work was supported by the Israeli Science
	Foundation, Grant No. 1233/13, and by the Wolfson foundation's High field Physics and attosecond science grant.


\begin{thebibliography}{10}
	
	\bibitem{brabec2000intense}
	Thomas Brabec and Ferenc Krausz.
	\newblock Intense few-cycle laser fields: Frontiers of nonlinear optics.
	\newblock {\em Reviews of Modern Physics}, 72(2):545, 2000.
	
	\bibitem{kapteyn2007harnessing}
	Henry Kapteyn, Oren Cohen, Ivan Christov, and Margaret Murnane.
	\newblock Harnessing attosecond science in the quest for coherent x-rays.
	\newblock {\em Science}, 317(5839):775--778, 2007.
	
	\bibitem{corkum2007attosecond}
	PB~Corkum and Ferenc Krausz.
	\newblock Attosecond science.
	\newblock {\em Nature Physics}, 3(6):381--387, 2007.
	
	\bibitem{popmintchev2012bright}
	Tenio Popmintchev, Ming-Chang Chen, Dimitar Popmintchev, Paul Arpin, Susannah
	Brown, Skirmantas Ali{\v{s}}auskas, Giedrius Andriukaitis, Tadas
	Bal{\v{c}}iunas, Oliver~D M{\"u}cke, Audrius Pugzlys, et~al.
	\newblock Bright coherent ultrahigh harmonics in the kev x-ray regime from
	mid-infrared femtosecond lasers.
	\newblock {\em science}, 336(6086):1287--1291, 2012.
	
	\bibitem{corkum1993plasma}
	Paul~B Corkum.
	\newblock Plasma perspective on strong field multiphoton ionization.
	\newblock {\em Physical Review Letters}, 71(13):1994, 1993.
	
	\bibitem{lewenstein1994theory}
	M~Lewenstein, Ph~Balcou, M~Yu Ivanov, Anne L’huillier, and Paul~B Corkum.
	\newblock Theory of high-harmonic generation by low-frequency laser fields.
	\newblock {\em Physical Review A}, 49(3):2117, 1994.
	
	\bibitem{durfee1999phase}
	Charles~G Durfee~III, Andy~R Rundquist, Sterling Backus, Catherine Herne,
	Margaret~M Murnane, and Henry~C Kapteyn.
	\newblock Phase matching of high-order harmonics in hollow waveguides.
	\newblock {\em Physical Review Letters}, 83(11):2187, 1999.
	
	\bibitem{popmintchev2008extended}
	Tenio Popmintchev, Ming-Chang Chen, Oren Cohen, Michael~E Grisham, Jorge~J
	Rocca, Margaret~M Murnane, and Henry~C Kapteyn.
	\newblock Extended phase matching of high harmonics driven by mid-infrared
	light.
	\newblock {\em Optics letters}, 33(18):2128--2130, 2008.
	
	\bibitem{popmintchev2009phase}
	Tenio Popmintchev, Ming-Chang Chen, Alon Bahabad, Michael Gerrity, Pavel
	Sidorenko, Oren Cohen, Ivan~P Christov, Margaret~M Murnane, and Henry~C
	Kapteyn.
	\newblock Phase matching of high harmonic generation in the soft and hard x-ray
	regions of the spectrum.
	\newblock {\em Proceedings of the National Academy of Sciences},
	106(26):10516--10521, 2009.
	
	\bibitem{lifshitz2005photonic}
	Ron Lifshitz, Ady Arie, and Alon Bahabad.
	\newblock Photonic quasicrystals for nonlinear optical frequency conversion.
	\newblock {\em Physical review letters}, 95(13):133901, 2005.
	
	\bibitem{armstrong1962interactions}
	JA~Armstrong, N~Bloembergen, J~Ducuing, and PS~Pershan.
	\newblock Interactions between light waves in a nonlinear dielectric.
	\newblock {\em Physical Review}, 127(6):1918, 1962.
	
	\bibitem{bahabad2010quasi}
	Alon Bahabad, Margaret~M Murnane, and Henry~C Kapteyn.
	\newblock Quasi-phase-matching of momentum and energy in nonlinear optical
	processes.
	\newblock {\em Nature Photonics}, 4(8):570--575, 2010.
	
	\bibitem{dromey2007quasi}
	B~Dromey, M~Zepf, M~Landreman, and SM~Hooker.
	\newblock Quasi-phasematching of harmonic generation via multimode beating in
	waveguides.
	\newblock {\em Optics express}, 15(13):7894--7900, 2007.
	
	\bibitem{ganeev2014quasi}
	RA~Ganeev, M~Suzuki, and H~Kuroda.
	\newblock Quasi-phase-matching of high-order harmonics in multiple plasma jets.
	\newblock {\em Physical Review A}, 89(3):033821, 2014.
	
	\bibitem{gibson2003coherent}
	Emily~A Gibson, Ariel Paul, Nick Wagner, David Gaudiosi, Sterling Backus,
	Ivan~P Christov, Andy Aquila, Eric~M Gullikson, David~T Attwood, Margaret~M
	Murnane, et~al.
	\newblock Coherent soft x-ray generation in the water window with quasi-phase
	matching.
	\newblock {\em Science}, 302(5642):95--98, 2003.
	
	\bibitem{pirri2008enhancing}
	Angela Pirri, Chiara Corsi, and Marco Bellini.
	\newblock Enhancing the yield of high-order harmonics with an array of gas
	jets.
	\newblock {\em Physical Review A}, 78(1):011801, 2008.
	
	\bibitem{paul2003quasi}
	A~Paul, RA~Bartels, R~Tobey, H~Green, S~Weiman, IP~Christov, MM~Murnane,
	HC~Kapteyn, and S~Backus.
	\newblock Quasi-phase-matched generation of coherent extreme-ultraviolet light.
	\newblock {\em Nature}, 421(6918):51--54, 2003.
	
	\bibitem{ganeev2015influence}
	Rashid~A Ganeev, Valer To{\c{s}}a, Katalin Kov{\'a}cs, Masayuki Suzuki, Shin
	Yoneya, and Hiroto Kuroda.
	\newblock Influence of ablated and tunneled electrons on quasi-phase-matched
	high-order-harmonic generation in laser-produced plasma.
	\newblock {\em Physical Review A}, 91(4):043823, 2015.
	
	\bibitem{kovacs2012quasi}
	Katalin Kov{\'a}cs, Emeric Balogh, J{\'a}nos Hebling, Valer To{\c{s}}a, and
	Katalin Varj{\'u}.
	\newblock Quasi-phase-matching high-harmonic radiation using chirped thz
	pulses.
	\newblock {\em Physical review letters}, 108(19):193903, 2012.
	
	\bibitem{serrat2010all}
	Carles Serrat and Jens Biegert.
	\newblock All-regions tunable high harmonic enhancement by a periodic static
	electric field.
	\newblock {\em Physical review letters}, 104(7):073901, 2010.
	
	\bibitem{zhang2007quasi}
	Xiaoshi Zhang, Amy~L Lytle, Tenio Popmintchev, Xibin Zhou, Henry~C Kapteyn,
	Margaret~M Murnane, and Oren Cohen.
	\newblock Quasi-phase-matching and quantum-path control of high-harmonic
	generation using counterpropagating light.
	\newblock {\em Nature Physics}, 3(4):270--275, 2007.
	
	\bibitem{o2014quasi}
	Kevin O’Keeffe, David~T Lloyd, and Simon~M Hooker.
	\newblock Quasi-phase-matched high-order harmonic generation using tunable
	pulse trains.
	\newblock {\em Optics express}, 22(7):7722--7732, 2014.
	
	\bibitem{faccio2010modulated}
	Daniele Faccio, Carles Serrat, Jos{\'e}~M Cela, Albert Farr{\'e}s, Paolo
	Di~Trapani, and Jens Biegert.
	\newblock Modulated phase matching and high-order harmonic enhancement mediated
	by the carrier-envelope phase.
	\newblock {\em Physical Review A}, 81(1):011803, 2010.
	
	\bibitem{peatross2004phase}
	Justin Peatross, Julia~R Miller, Kelly~R Smith, Steven~E Rhynard, and
	Benjamin~W Pratt.
	\newblock Phase matching of high-order harmonic generation in helium-and
	neon-filled gas cells.
	\newblock {\em Journal of Modern Optics}, 51(16-18):2675--2683, 2004.
	
	\bibitem{rundquist1998phase}
	Andy Rundquist, Charles~G Durfee, Zenghu Chang, Catherine Herne, Sterling
	Backus, Margaret~M Murnane, and Henry~C Kapteyn.
	\newblock Phase-matched generation of coherent soft x-rays.
	\newblock {\em Science}, 280(5368):1412--1415, 1998.
	
	\bibitem{auguste2007quasi}
	T~Auguste, B~Carr{\'e}, and P~Sali{\`e}res.
	\newblock Quasi-phase-matching of high-order harmonics using a modulated atomic
	density.
	\newblock {\em Physical Review A}, 76(1):011802, 2007.
	
	\bibitem{balcou1997generalized}
	Philippe Balcou, Pascal Salieres, Anne L'Huillier, and Maciej Lewenstein.
	\newblock Generalized phase-matching conditions for high harmonics: The role of
	field-gradient forces.
	\newblock {\em Physical Review A}, 55(4):3204, 1997.
	
	\bibitem{lewenstein1995phase}
	Maciej Lewenstein, Pascal Salieres, and Anne L’huillier.
	\newblock Phase of the atomic polarization in high-order harmonic generation.
	\newblock {\em Physical Review A}, 52(6):4747, 1995.
	
	\bibitem{mcgloin2005bessel}
	D~McGloin and K~Dholakia.
	\newblock Bessel beams: diffraction in a new light.
	\newblock {\em Contemporary Physics}, 46(1):15--28, 2005.
	
	\bibitem{diskin2015phase}
	Tzvi Diskin, Ofer Kfir, Avner Fleischer, and Oren Cohen.
	\newblock Phase modulation in polarization beating quasi-phase-matching of
	high-order-harmonic generation.
	\newblock {\em Physical Review A}, 92(3):033807, 2015.
	
	\bibitem{hadas2016periodic}
	Itai Hadas and Alon Bahabad.
	\newblock Periodic density modulation for quasi-phase-matching of optical
	frequency conversion is inefficient under shallow focusing and constant
	ambient pressure.
	\newblock {\em Optics Letters}, 41(17):4000--4003, 2016.
	
	\bibitem{ruffner2012optical}
	David~B Ruffner and David~G Grier.
	\newblock Optical conveyors: a class of active tractor beams.
	\newblock {\em Physical review letters}, 109(16):163903, 2012.
	
	\bibitem{boyd2003nonlinear}
	Robert~W Boyd.
	\newblock {\em Nonlinear optics}.
	\newblock Academic press, 2003.
	
	\bibitem{sandhu2006generation}
	Arvinder~S Sandhu, Etienne Gagnon, Ariel Paul, Isabell Thomann, Amy Lytle,
	Tracy Keep, Margaret~M Murnane, Henry~C Kapteyn, and Ivan~P Christov.
	\newblock Generation of sub-optical-cycle,
	carrier-envelope-phase—insensitive, extreme-uv pulses via nonlinear
	stabilization in a waveguide.
	\newblock {\em Physical Review A}, 74(6):061803, 2006.
	
	\bibitem{thomann2009characterizing}
	I~Thomann, A~Bahabad, X~Liu, R~Trebino, MM~Murnane, and HC~Kapteyn.
	\newblock Characterizing isolated attosecond pulses from hollow-core waveguides
	using multi-cycle driving pulses.
	\newblock {\em Optics express}, 17(6):4611--4633, 2009.
	
	\bibitem{bahabad2011manipulating}
	Alon Bahabad, Margaret~M Murnane, and Henry~C Kapteyn.
	\newblock Manipulating nonlinear optical processes with accelerating light
	beams.
	\newblock {\em Physical Review A}, 84(3):033819, 2011.
	
	\bibitem{fleischer2014spin}
	Avner Fleischer, Ofer Kfir, Tzvi Diskin, Pavel Sidorenko, and Oren Cohen.
	\newblock Spin angular momentum and tunable polarization in high-harmonic
	generation.
	\newblock {\em Nature Photonics}, 8(7):543--549, 2014.
	
	\bibitem{gariepy2014creating}
	Genevieve Gariepy, Jonathan Leach, Kyung~Taec Kim, Thomas~J Hammond, Eugene
	Frumker, Robert~W Boyd, and Paul~B Corkum.
	\newblock Creating high-harmonic beams with controlled orbital angular
	momentum.
	\newblock {\em Physical review letters}, 113(15):153901, 2014.
	
	\bibitem{vieira2016high}
	J~Vieira, RMGM Trines, EP~Alves, RA~Fonseca, JT~Mendon{\c{c}}a, R~Bingham,
	P~Norreys, and LO~Silva.
	\newblock High orbital angular momentum harmonic generation.
	\newblock {\em Physical Review Letters}, 117(26):265001, 2016.
	
	\bibitem{bahabad2008quasi}
	Alon Bahabad, Oren Cohen, Margaret~M Murnane, and Henry~C Kapteyn.
	\newblock Quasi-phase-matching and dispersion characterization of harmonic
	generation in the perturbative regime using counterpropagating beams.
	\newblock {\em Optics express}, 16(20):15923--15931, 2008.
	
	\bibitem{weiner2011ultrafast}
	Andrew Weiner.
	\newblock {\em Ultrafast optics}, volume~72.
	\newblock John Wiley \& Sons, 2011.
	
	\bibitem{sharipov2004numerical}
	Felix Sharipov.
	\newblock Numerical simulation of rarefied gas flow through a thin orifice.
	\newblock {\em Journal of Fluid Mechanics}, 518:35--60, 2004.
	
	\bibitem{l1991higher}
	A~L’Huillier, KJ~Schafer, and KC~Kulander.
	\newblock Higher-order harmonic generation in xenon at 1064 nm: The role of
	phase matching.
	\newblock {\em Physical review letters}, 66(17):2200, 1991.
	
	\bibitem{gaarde2008macroscopic}
	Mette~B Gaarde, Jennifer~L Tate, and Kenneth~J Schafer.
	\newblock Macroscopic aspects of attosecond pulse generation.
	\newblock {\em Journal of Physics B: Atomic, Molecular and Optical Physics},
	41(13):132001, 2008.
	
	\bibitem{rego2016nonperturbative}
	Laura Rego, Julio San~Rom{\'a}n, Antonio Pic{\'o}n, Luis Plaja, and Carlos
	Hern{\'a}ndez-Garc{\'\i}a.
	\newblock Nonperturbative twist in the generation of extreme-ultraviolet vortex
	beams.
	\newblock {\em Physical Review Letters}, 117(16):163202, 2016.
	
\end{thebibliography}
\end{document}